\documentclass[doublecol]{epl2} 

\usepackage[utf8]{inputenc} 
\usepackage{epsfig}
\usepackage{amsmath}
\usepackage{subfigure}
\usepackage{epstopdf} 
\usepackage{graphicx}
\usepackage{color}
\usepackage{pict2e}
\usepackage{braket}

\title{Contextuality and correlation loopholes are equivalent}

\author{\'Alvaro G. L\'{o}pez \inst{1}}

\institute{Nonlinear Dynamics, Chaos and Complex Systems Group, Departamento de F\'{i}sica, Universidad Rey Juan Carlos, Tulip\'{a}n s/n, 28933 M\'{o}stoles, Madrid, Spain \inst{1} \\ \textnormal{alvaro.lopez@urjc.es}}

\abstract{We show that contextual hidden variables including the effect of the measuring devices can be backward-propagated by means of the Green's function to initial Cauchy hidden data. If this data is uncorrelated in spacelike-disjoint sets, the CHSH-Bell inequality can be derived. However, the correlation loophole remains unclosed.}

\begin{document}
\maketitle

\section{Introduction}
Since the original demonstration of the Bell-Kochen-Specker theorem, contextuality has systematically proven to be an essential feature of quantum mechanical systems \cite{how14}. During the last decade, the idea that hidden variable models incorporating the effects of the measuring apparatus might be indispensable to understand the violation of Bell-type inequalities has also gained increased support \cite{mar15,mar17}. 

Contrary to local realism, the contextual paradigm claims that the experimental outcomes cannot simply be thought of as revealing pre-existing values of a quantum particle. Instead, it is proposed that they emerge from their interaction with the apparatus. The contextuality loophole has not yet been closed, setting a moratorium on the validity of Bell tests \cite{nie11,lar14}. 

Among contextual advocates, some authors have pointed out the crucial role of hidden electrodynamic zero-point fluctuations \cite{nie11}. This line of reasoning has been recently reinforced by studies with extended electrodynamic bodies. Retarded potentials lead to time-delayed self-interactions, which produce limit cycle oscillations through a Hopf bifurcation \cite{lop20,lop202,lop23}. The resulting jittery motion has a frequency closely related to the frequency of $\emph{zitterbewegung}$ and can be identified as a particular instance of self-oscillation \cite{jen13}. In analogy to experiments with walking silicone oil droplets \cite{cou05}, it has been proposed that this nonlinear oscillation is responsible for the wave-particle duality of quantum particles \cite{lop20}.

The mathematical proof of Bell-type theorems, from their original derivations \cite{bel64,cla69} to the most general case provided to present date \cite{mor06}, relies on the fact that the hidden variables on which the probability measure depends are independent of the measurement process. At the time of detection, the electromagnetic interaction between the particle and the measuring device introduces a dependence of the probability density on its orientation, precluding the derivation of Bell-type inequalities. 

In classical field theories, this functional dependence also appears when initial hidden fields are used to express the spin-correlation integrals. The probability density can be expressed as a function of hidden fields with support on a compact domain in the Cauchy hypersurface enclosing the initial data, by means of the Green's function of the dynamical theory. This entails a reduction of the contextuality loophole to a correlation loophole. The later was overlooked by Bell \cite{bel04} (and also in a reply to his theory of local beables \cite{shi76}) as a possible source of violation of the CHSH-Bell inequality in tests of local realism \cite{mor06}.

\section{The Einstein-Podolsky-Rosen-Bohm experiment} 

We consider a classical experiment proposed by Bohm and Aharonov \cite{boh57,hen15}. A similar argument for optical quantum systems \cite{asp82} used to test the locality of quantum mechanics and its related field theories, is straightforward to implement.

From the point of view of a local classical field theory ($e.g.$ Einstein-Maxwell electrodynamics), the \emph{internal} angular momentum could be related to the magnetic moment of an electromagnetic topological soliton \cite{mis57,fab01}. Alternatively, we can consider that particles are localized point charge distributions evolving under the action of electrodynamic retarded fields, with their internal properties (charge, mass and spin) embedded in the point. No specific hypothesis about the nature of the particles, and how the spin emerges from the currents inside them, are essential to our argument. The only requirement to prove our main result is the existence and uniqueness of solutions to the partial differential equations describing the dynamical evolution of the fields, once the initial field configuration and the boundary conditions have been specified \cite{mis57}.

Before entering the Stern-Gerlach (S-G) apparatus, the magnetic moment of each element of the entangled pair could be dynamically evolving in a rather complicated fashion inside a sphere. The process of measurement involves the orientation of its intrinsic magnetic moment along the external magnetic field, from the moment that each particle enters the S-G apparatus, to the first instant of time when its spin has fully aligned with the external non-uniform magnetic field.

\section{Results}
 
As usual, we now define $A_a[\lambda_A(x_a,T)]$ and $B_b[\lambda_B(x_b,T)]$ as the two observables representing the result of measuring the spin of each particle along the directions $a$ and $b$ in the Euclidean space, with outcomes of either $1$ for spin up, or $-1$ for spin down. Then, the spin-correlation integral of the entangled pair at the time of measurement $t=T$ can be written as 
\begin{equation}
C_{a b}(T)=\int_\mathcal{C} A_a[\lambda_A(x_a,T)] B_b[\lambda_B(x_b,T)] p_{a b}[\lambda,T] \mathcal{D}\lambda,
\end{equation}
with $\lambda=(\lambda_A(x_a,T),\lambda_B(x_b,T))$ the value of the hidden fields inside the particles, whose centers of mass are located inside a small region, described by $x_a$ and $x_b$ respectively, where most of the particle's energy is stored, once the spin alignment has completed. The sample space of all possible field configurations with compact support in the region where the particles are placed has been denoted as $\mathcal{C}$. Recall, because of contextuality, the probability density of the random hidden fields at the time of measurement $p_{a b}[\lambda,T]$ depends on the orientations of the magnetic fields of the two S-G artifacts. 

\begin{figure}
\centering
\includegraphics[width=0.45\textwidth]{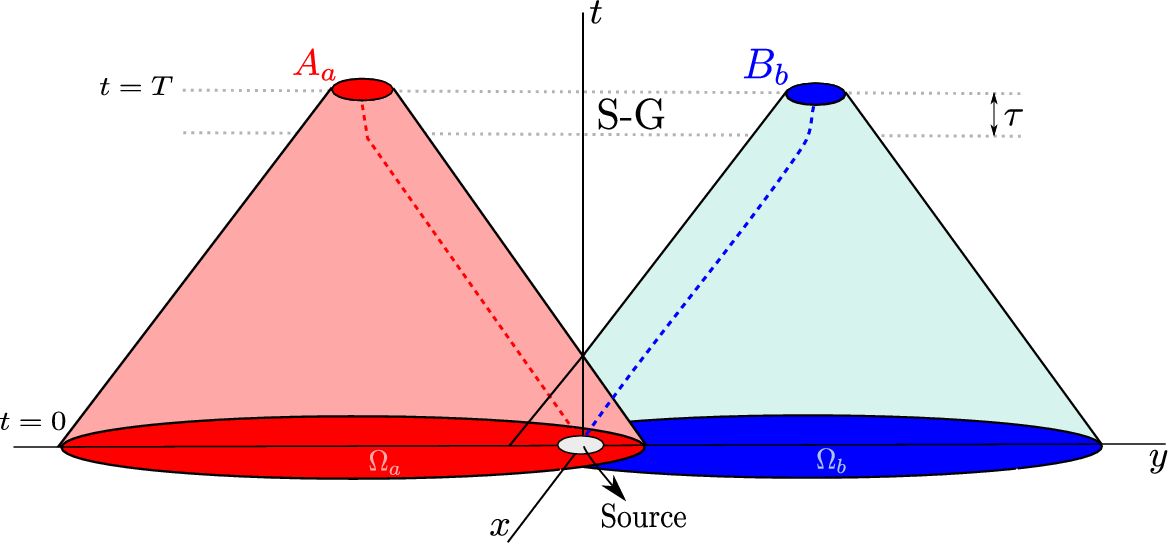}
\caption{Minkowski diagram of the EPRB experiment in 2+1 dimensions. Two entangled particles are emitted at the source at time $t=0$. Then, they follow their world lines until they enter their respective apparatuses at $t=\tau$. At $t=T$ these two particles have aligned their intrinsic magnetic moments with the external magnetic field. The causally connected initial domains in the initial Cauchy surface at time $t=0$ for the observers $A_a$ and $B_b$ are represented by $\Omega_a$ and $\Omega_b$.} \label{fig:1}
\end{figure}

To revert the flow to the initial setting, we have to take into account the light cones of the particles, which connect their field configuration at the time of measurement with the fields inside an initial Cauchy slice of their causal past at $t=0$, as depicted in Fig.~\ref{fig:1}. From the point of view of the partial differential equations that govern the dynamics of the fields \cite{gre00}, the Green’s function $G(x',x,t)$ selecting the causal region of the initial data $\lambda_0(x)$ on which the present hidden fields depend, satisfies the identity $\lambda_a(x',t)=\int G(x',x,t)\lambda_0(x)d^3x$.

Therefore, the correlation of spins, when expressed in terms of the initial hidden fields, is given by the functional integral
\begin{equation}
C_{a b}=\int_{\mathcal{C}_{0}} \hat{A}_a[\lambda_0(x)] \hat{B}_b[\lambda_0(x)] p_{ab}[\lambda_0(x)]  \mathcal{D}\lambda_0,
\label{eq:2}
\end{equation}
where $\hat{A}_a[\lambda_0(x)] \equiv A_a[\int G(x',x,t)\lambda_0(x)d^3x]$ has been defined as a functional, and similarly for $\hat{B}_b[\lambda_0(x)]$. The probability densities are related through a change of variables, in the form $p_{a b}[\lambda,t]=\int_{\mathcal{C}_{0}} \delta[\lambda_A(x_a,t)-\lambda_a(x_a,t)]\delta[\lambda_B(x_b,t)-\lambda_B(x_b,t)] p_{ab}[\lambda_0(x)] \mathcal{D}\lambda_0$. 

The Eq.~\eqref{eq:2} is defined over the set of functions $\mathcal{C}_0$ having compact support in the spatial domain $\Sigma=\Omega_{a} \cup \Omega_{b}$, enclosed in the Cauchy hypersurface at $t=0$. If the initial fields are only correlated in the domain $\Sigma_{c}=\Omega_{a} \cap \Omega_{b}$ depicted in Fig.~\ref{fig:1}, the CHSH-Bell inequality can be immediately derived. We only have to average over the regions $\Sigma_{a}=\Omega_{a}/\Sigma_c$ and $\Sigma_{b}=\Omega_{b}/\Sigma_c$, to get rid of the contextual hidden field fluctuations. 

For clarity, but without loss of generality, we give the proof assuming that the fields are uncorrelated in any two spacelike-separated sets in the initial Cauchy surface. Nevertheless, it suffices to consider that fields in $\Sigma_a$ and $\Sigma_b$ are mutually uncorrelated and that the fields in $\Sigma_{c}$ are also independent of $a$ and $b$. This conspiracy also involves field correlations and is thoroughly discussed in Refs.~\cite{mor06,shi76} and has been considered as particular case of superdeterminism \cite{hos20}. 

Under such hypotheses, the probability density $p_{ab}[\lambda_0(x)]$ can be expressed as a product of densities $p_{a}[\lambda_a(x)]p_{b}[\lambda_b(x)]p[\lambda_c(x)]$, where $\lambda_a(x)=\{\lambda_{0}(x):x\in \Sigma_{a}\}$, $\lambda_b(x)=\{\lambda_{0}(x):x \in \Sigma_{b}\}$ and $\lambda_c(x)=\{\lambda_{0}(x):x\in \Sigma_{c}\}$. Now we write the functions $\hat{A}_{a}[\lambda_a(x),\lambda_c(x)]$ and $\hat{B}_{b}[\lambda_b(x),\lambda_c(x)]$, as well. This yields the functional integral  
\begin{equation}
\begin{split}
C_{ab} = \iiint  & \hat{A}_a[\lambda_a(x),\lambda_c(x)] \hat{B}_b[\lambda_b(x),\lambda_c(x)]  \\
 & p_{a}[\lambda_a(x)] p_{b}[\lambda_b(x)] p[\lambda_c(x)] \mathcal{D} \lambda_a  \mathcal{D} \lambda_b  \mathcal{D} \lambda_c.
\end{split}
\end{equation}
After averaging out the fluctuations $\bar{A}_{a}[\lambda_c(x)]=\int \hat{A}_a[\lambda_a(x),\lambda_c(x)] p_{a}[\lambda_a(x)] \mathcal{D} \lambda_a$, and the same for $\bar{B}_{b}[\lambda_c(x)]$, the following Bell-type integral results
\begin{equation}
C_{a b}=\int \bar{A}_a[\lambda_c(x)] \bar{B}_b[\lambda_c(x)] p[\lambda_c(x)]  \mathcal{D}\lambda_c.
\label{eq:4}
\end{equation}
It has been pointed out that it is impossible to accomplish these experiments \cite{mar15,mar17}. This fact is irrelevant, since Eq.~\eqref{eq:4} allows to derive the CHSH-Bell inequality
\begin{equation}
|C_{a b}-C_{a b'}+C_{a' b}+C_{a' b'}| \leq 2.
\label{eq:5}
\end{equation}

Therefore, the Eq.~\eqref{eq:4} must describe the same type of correlations. Of note, the existence of Eq.~\eqref{eq:4} is a necessary and sufficient condition for the fulfillment of Eq.~\eqref{eq:5} \cite{fine82}. 

\section{Last instant and free choices}

We can also extend the previous argument to systems in which the orientations of the S-G are set right before the measurements are made \cite{asp82}. The dependence of the probability density $p_{ab}[\lambda_0 (x)]$ on the orientations $a$ and $b$ cannot be avoided in general. To compute the correlation when we select a definite orientation ($e.g.$, $a$) of the apparatus, we must disregard all the initial field configurations that evolve towards a different orientation ($e.g.$, $a'$). In the particular case of purely deterministic local field theories this is unavoidable, because the set of initial conditions in the Cauchy surface that lead to a specific orientation of the apparatus are always different from those leading to another orientation.

Concerning stochastic hidden variable models \cite{mor06}, we can represent random fluctuations by incorporating Langevin currents in the partial differential equations describing the field theory. As an example, we can consider a case where there exists a deterministic drift term in the Langevin stochastic differential equation governing the evolution of the dynamical fields, and the intensity of the random fluctuations is bounded. Then, it can happen that a particular orientation of the apparatus is not accessible through the stochastic flow from some hidden field configurations in the initial Cauchy surface, which would evolve in time towards some other orientation. Thus a restriction of events in the sample space of the initial hidden fields is imposed even when random dynamical settings are used.

In recent works, the dependence of the probability density $p_{ab}[\lambda_{0}(x)]$ on the orientation of the apparatuses has been considered a revisited type of superdeterminism \cite{hos20}. However, in the present case the correlations involve fields taking values at two points in $\Sigma_a$ and $\Sigma_b$, but not necessarily those involving the apparatuses and their orientations. The existence of $p[\lambda_{c}(x)]$ also assumes that the orientation of the S-G and the hidden fields at the source are independent. Moreover, since our argument is extensible to models described by stochastic partial differential equations, we do not need to assume determinism. It is evident that the correlation loophole is different from superdeterminism, specially as this conspiracy is frequently considered \cite{mor06,hos20}, and as was discussed by Bell initially \cite{bel042}.

\section{Conclusions} 

The simplest proofs of the CHSH-Bell inequality rely on the existence of a joint probability density including experimental realizations in which the same S-G apparatus has two different orientations \cite{fine82}. The non-Kolmogorovness of such a probability density \cite{khr09} is sometimes attributed to the contextual role of the measuring devices \cite{mar17}. 

We have shown that contextual implications can be circumvented by reverting the flow to the initial hidden fields, when the particles are created at their source. Instead, we have to prove that the electrodynamic field fluctuations are uncorrelated in the spacelike-separated sets $\Sigma_a$ and $\Sigma_b$. As has been recently shown using silicone droplet models, particles communicating through correlated background fields can synchronize their dynamics and violate the CHSH-Bell inequality \cite{kon22}. However, these experiments use static analyzers and cannot close Bell's locality loophole.

This correlation loophole (likewise, the contextual) is far more feasible to close from an empirical point of view. A general experimental protocol to overcome the independence of space-like separated sets must maximize the size of $\Sigma_c$, excluding at the same time the apparatuses from it. Such tests are favored by our analysis to avoid a statistical dependence between $\Sigma_a$ and $\Sigma_b$.

For example, in the case that two entangled bodies recede each other in opposite directions with uniform motion, the radius of $\Sigma_c$ is equal to $D(1-\beta)/\beta$, with $D$ the distance between the measuring device and the source, and $\beta$ the speed of the particles relative to the speed of light in the laboratory frame. This equation suggests that the optimum speed for two massive bodies would be one above, but as close as possible, half the speed of light. 

On the other hand, experiments with correlated photon pairs receding from each other at the speed of light should be disregarded as an appropriate experimental environment to discard contextual theories like classical electrodynamics and its covariant generalization, since $\Sigma_c$ is reduced to its minimum in these experiments with massless particles. Counterintuitively, apart from closing the locality loophole, last-instant choices are not particularly favored by our result, as well. If $a$ and $b$ are dynamically set at some time right before the particles enter the S-G devices, a correlation between the fields in $\Sigma_c$ and each orientation of the measuring instruments must be prevented \cite{fle22}. Otherwise, the probability density $p[\lambda_c(x)]$ could inherit a dependence on $a$ and $b$, hindering the warrant of the no-conspiracy previously referred and precluding the derivation of Eq.~\eqref{eq:5} \cite{mor06,shi76}.


\begin{thebibliography}{0}
\bibitem{gre00}
The existence of a Green's function holds for linear theories only. It is here assumed for aesthetic purposes. In the case of nonlinear theories the argument is similar, but using the existence and uniqueness of solutions to the action of the field theory Lagrangian, instead.
\end{thebibliography}
\end{document}